\documentclass[aps,prb,twocolumn,showpacs,superscriptaddress,floatfix]{revtex4}
\usepackage{graphicx}
\usepackage{multirow}
\usepackage{color}
\usepackage{mciteplus}
\begin{document}
\title{Stability and electronic properties of two-dimensional silicene and germanene on graphene}
\author{Yongmao Cai}
\affiliation{Institute of Atomic and Molecular Sciences, Academia Sinica, Taipei 10617, Taiwan}
\author{Chih-Piao Chuu}
\email{taimosa@gmail.com}
\affiliation{Institute of Atomic and Molecular Sciences,  Academia Sinica, Taipei 10617, Taiwan}
\author{C. M. Wei}
\affiliation{Institute of Atomic and Molecular Sciences,  Academia Sinica, Taipei 10617, Taiwan}
\author{M. Y. Chou}
\email{meiyin.chou@physics.gatech.edu}
\affiliation{Institute of Atomic and Molecular Sciences,  Academia Sinica, Taipei 10617, Taiwan}
\affiliation{Department of Physics, National Taiwan University, Taipei, Taiwan}
\affiliation{School of Physics, Georgia Institute of Technology, Atlanta, Georgia 30332, USA}
\date{\today}
\begin{abstract}
We present first-principles calculations of silicene/graphene and germanene/graphene bilayers. Various supercell models are constructed in the calculations in order to reduce the strain of the lattice-mismatched bilayer systems. Our energetics analysis and electronic structure results suggest that graphene can be used as a substrate to synthesize monolayer silicene and germanene. Multiple phases of single crystalline silicene and germanene with different orientations relative to the substrate could coexist at room temperature. The weak interaction between the overlayer and the substrate preserves the low-buckled structure of silicene and germanene, as well as their linear energy bands. The gap induced by breaking the sublattice symmetry in silicene on graphene can be up to 57 meV.
\end{abstract}
\maketitle
\graphicspath{{figures/}}
\section{Introduction}
Silicene, a two-dimensional (2D) monolayer of silicon, consists of a honeycomb lattice of atoms with a buckled configuration.\cite{ciraci09} Electronically, a linear dispersion in the vicinity of Dirac points gives rise the feature of massless Dirac fermions. The band-gap engineering in silicene can be accomplished by electrical means,\cite{ni12,drummond12,Ezawa12NJP} and the interplay between the non-negligible spin-orbit coupling (SOC) and electromagnetic (EM) field can be used to probe the physics of the quantum phase transition; novel quantum phenomena such as the quantum spin Hall (QSH)\cite{kane05qsh} effect and the quantum anomalous Hall (QAH)\cite{yu10qah} effect are expected to be observed in this promising new 2D material.\cite{yao11,ezawa12qah} Considerable research efforts have been reported for the synthesis of silicene on various substrates: for example, the formation of epitaxial silicene on Ag(111),\cite{lay12,feng12,chiappe12,enriquez12,Jamgotchian12,Ezawa13epjb} (0001)-oriented zirconium diboride on Si(111) wafers,\cite{fleurence12} and Ir(111)\cite{meng13} has been reported recently. However, the complicated surface reconstructions of silicene originated from the interaction with the substrate have been observed on the Ag(111) surface, giving rise to structures deviating from the low-buckled (LB) honeycomb configuration predicted for the freestanding monolayer.\cite{ciraci09} As a result, the existence of Dirac fermions in silicene is under debate due to the lack of direct evidences from experiment.\cite{lin13,Rubio13}

\begin{figure}
\includegraphics[width=2.9in]{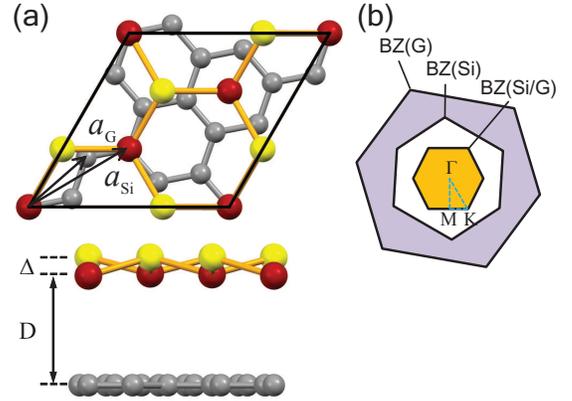}
\caption{(Color online) (a) Top and side views of the atomic structure of Si($\sqrt{3}$)/G($\sqrt{7}$) (see text for the definition of the supercell notation), where red and yellow atoms represent the two sublattices of Si atoms separated by $\Delta$ = 0.62 $\text{\AA}$ vertically, and the grey spheres represent carbon atoms in the graphene layer at a separation of $D$=3.3 $\text{\AA}$ from the lower silicon layer. The lattice vectors $\vec{a}_{\text{Si}}$ and $\vec{a}_{\text{G}}$ of the (1$\times$1) unit cell of silicene and graphene, respectively, have a relative rotational angel of $10.9^\circ$ between them. (b) The first Brillouin zones of Si($\sqrt{3}$)/G($\sqrt{7}$) (orange), 1$\times$1 silicene (white), and 1$\times$1 graphene (purple) are plotted.
The blue triangle indicates the path $\Gamma$MK of the band structure.}
\end{figure}

The aforementioned difficulties prompt us to consider weakly interacting substrates that may preserve the symmetrically buckled structure of 2D silicene and therefore its linear energy dispersion. In this paper, we propose the possibility of forming a bilayer structure, silicene on graphene (Si/G), as a path to grow silicene on the isostructural and weakly interactive graphene substrate. We find from our calculations that the Si/G bilayer is locally stable with no imaginary phonon frequencies, the electronic band structure barely changes with the linear dispersion preserved in the vicinity of the Dirac points, and a slight electron transfer (about 2$\times10^{12}$ cm$^{-2}$) from silicene to graphene occurs. Even without including the SOC, a gap of up to 57 meV can be opened by breaking the sublattice symmetry in the bilayer. We have performed a similar analysis for the bilayer germanene/graphene (Ge/G). Two-dimensional germanene has a similar LB honeycomb structure as silicene, except that its SOC effect is about ten times larger than that in silicene.\cite{yao11} For both Si/G and Ge/G, our result suggests that multiple phases with different rotational angles could coexist at room temperature.

\begin{figure}
\includegraphics[width=2.9in]{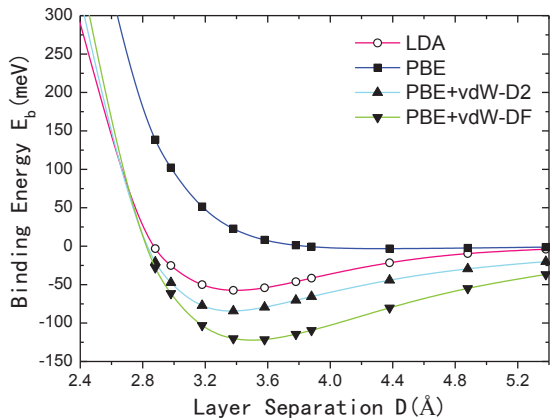}
\caption{ (Color online). Interlayer binding energy per Si atom of the Si($\sqrt{7}$)/G(4) bilayer as a function of interlayer spacing. Results using different exchange-correlation functionals are show. See text for the geometry and the binding energy definition.}
\end{figure}

The paper is organized as following. The computational methodology of this first-principles calculation is described in next section. The stability of bilayer Si/G and Ge/G is analyzed in Sec. III, followed by a discussion of its electronic properties. Finally, we provide summaries of the study.

\section{Computational methodology}

\begin{table}[]\caption{Supercell models for the Si/G and Ge/G bilayers where G stands for graphene. Each model is created by combining different supercells in individual layers as indicated, with a rotational angle $\phi$ between the two layers.  $L_{\text{Si}}$, $L_{\text{G}}$, $L_{\text{Si/G}}$, and $L_{\text{Ge/G}}$ are the LDA lattice constants for the particular supercells of silicene, graphene, Si/G, and Ge/G, respectively, while $a$ is the effective lattice constant of silicene (germanene) in the relaxed bilayer structure. $\theta$ and $\Delta$ are the bond angle and buckling distance in silicene (or germanene), respectively. $d$ is the strain in silicene or germanene as defined in the text.}
\begin{tabular}{c c c c c c c c c c c}
\hline
Si/G &  $a$($\text{\AA}$)&$\phi$($^\circ$)&$\theta(^\circ)$&$\Delta$($\text{\AA}$)&$L_{\text{\text{Si}}}$&$L_{\text{G}}$& $L_{\text{Si/G}}$&$d$($\%$)\\ [0.5ex]
\hline
$\sqrt{7}$/4 &3.73 &19.1&112.2&0.64&10.13 &9.80 & 9.86&-2.7 \\[1ex]
$\sqrt{3}$/$\sqrt{7}$ &3.75 & 10.9&112.7&0.62&6.63 &6.47 &6.50& -2.0 \\[1ex]
$\sqrt{13}$/$\sqrt{31}$ &3.79 & 4.9&113.7&0.58&13.81 &13.61 &13.65& -1.2 \\[1ex]
4/$\sqrt{39}$ &3.82 & 16.1&114.7&0.53&15.32 &15.27 &15.28& -0.2 \\[1ex]
$\sqrt{21}$/$2\sqrt{13}$ &3.85 &  3&115.4&0.50&17.55 & 17.63 &17.63&0.5\\[1ex]
$\sqrt{19}$/$4\sqrt{3}$ &3.88 & 6.6&116.1&0.46&16.69 &16.94 &16.90&1.3 \\[1ex]
\hline
Ge/G & $a$($\text{\AA}$)&$\phi$($^\circ$)&$\theta(^\circ)$&$\Delta$($\text{\AA}$)&$L_{\text{\text{Ge}}}$&$L_{\text{G}}$& $L_{\text{Ge/G}}$&$d$($\%$)\\ [0.5ex]
\hline
$3\sqrt{3}$/$\sqrt{67}$ &3.87& 17.8&110.1&0.76&20.63 &20.02&20.13& -2.5 \\[1ex]
5/8 &3.93&0&111.2&0.72&19.85 &19.56 &19.62& -1.1 \\[1ex]
$2\sqrt{3}$/$\sqrt{31}$ &3.94 & 21&111.5&0.71&13.75& 13.61&13.66& -0.7 \\[1ex]
$\sqrt{39}$/$\sqrt{103}$ &3.98 & 8.4&112.1&0.69&24.79 &  24.82&24.83& 0.1 \\[1ex]
$\sqrt{7}$/$\sqrt{19}$ &4.03 & 4.3&113.2&0.64&10.50 & 10.66&10.65& 1.4 \\[1ex]
3/5 &4.06 & 0&113.8&0.62&11.91&12.23 &12.18&2.2 \\[1ex]
\hline\hline
\end{tabular}
\end{table}

First-principles calculations are performed based on density functional theory (DFT)\cite{dft1,dft2} using the Vienna $ab$ $initio$ simulation package (VASP).\cite{vasp1,vasp2} Valence wave functions are treated by the projector augmented wave (PAW) method\cite{paw} that uses pseudopotential operators but keeps the full all-electron wave functions. The interlayer interaction is checked by various exchange-correlation energy functionals, including the local density approximation (LDA)\cite{lda}, the Perdew-Burke-Ernzerhof (PBE) generalized gradient approximation,\cite{pbe} and the PBE with vdW corrections incorporated at two different levels: the vdW-D2 and vdW-DF functionals.\cite{vdw,sevdw}  The plane-wave energy cut-off is at least 400 eV. We have checked the convergence of $k$ points and used meshes containing at least 252 points in the primitive Brillouin zone (BZ) of graphene. A vacuum of 20 $\textrm{\AA}$ is used to eliminate the spurious interaction.  For structural optimization, all atoms are relaxed until the change of the energy and the force reaches 10$^{-5}$ eV or 10$^{-6}$ eV per cell and 10$^{-2}$ eV/$\textrm{\AA}$, respectively.

\section{Energetics of Silicene and Germanene on Graphene}

We first consider the energetics of the bilayer system of silicene or germanene on graphene. For monolayer graphene, silicene, and germanene, the lattice constants obtained from LDA are 2.45, 3.82, and 3.97~$\text{\AA}$, which agree well with previous published results.\cite{ciraci09} Given the lattice mismatch, we need to identify appropriate supercells in the calculations of the bilayer system by rotating the silicene (or germanene) layer with respect to the graphene substrate. For a 2D hexagonal lattice, it is possible to find supercells defined by longer lattice vectors at various angles from the primitive one. For example, the angles associated with the lattice vectors for the $\sqrt{3}\times\sqrt{3}$, $\sqrt{7}\times\sqrt{7}$, $\sqrt{13}\times\sqrt{13}$, $\sqrt{19}\times\sqrt{19}$, $\sqrt{21}\times\sqrt{21}$, $\sqrt{31}\times\sqrt{31}$, $\sqrt{39}\times\sqrt{39}$, $\sqrt{67}\times\sqrt{67}$, and $\sqrt{103}\times\sqrt{103}$ unit cells are 30$^\circ$, 19.1$^\circ$, 13.9$^\circ$, 23.4$^\circ$, 10.9$^\circ$, 9.0$^\circ$, 16.1$^\circ$, 12.2$^\circ$, and 24.5$^\circ$, respectively. By making different combinations of the supercells of silicene (or germanene) and graphene, one can construct bilayer systems with a small amount of strain.

The supercells we have considered along with their structural parameters are listed in Table I. For example, $\sqrt{3}$/$\sqrt{7}$ for silicon on graphene (Si/G) corresponds to a supercell consisting of $\sqrt{3}\times\sqrt{3}$ silicene and $\sqrt{7}\times\sqrt{7}$ graphene combined after a rotation of angle $\phi$ equal to $30^\circ - 19.1^\circ = 10.9^\circ$. This supercell configuration will be represented as Si($\sqrt{3}$)/G($\sqrt{7}$) in the text.

Figure 1 shows the atomic structure and the BZ of this bilayer system. The red and yellow spheres represent Si atoms in different layers of the buckled structure. An optimized structure is obtained when a Si atom in the lower layer is placed on top of a C atom in graphene. The buckling distance $\Delta$ is found to be 0.62 $\textrm{\AA}$ in this system, and the distance from graphene to the lower Si layer is 3.3 $\textrm{\AA}$ based on the LDA calculation (to be discussed below), indicating that it belongs to the class of van der Waals (vdW) type of heterostructures.\cite{geim13}

$L_{\text{Si/G}}$ ($L_{\text{Ge/G}}$) in Table I is the supercell length of the fully relaxed Si/G (Ge/G) bilayer determined by the LDA, while $L_{\text{Si}}$ ($L_{\text{Ge}}$)and $L_{\text{G}}$ are the lengths of corresponding supercells for monolayer silicene (germanene) and graphene, respectively. Note that $L_{\text{Si/G}}$ and $L_{\text{Ge/G}}$ are both closer to $L_{\text{G}}$, indicating that the strain in graphene is smaller than that in the silicene or germanene layer. This is expected since with a stronger $\sigma$ bond the energy cost for changing 1$\%$ of the graphene lattice constant is 7 meV per C atom, larger than the corresponding value of 3$\sim$4 meV per atom in silicene and germanene. The strain in the silicene layer is defined by
\begin{eqnarray}
\textit{d}=\frac{\textit{a}-\textit{a}_0}{\textit{a}_0}=\frac{L_{\text{Si/G}}-L_{\text{Si}}}{L_{\text{Si}}},
\end{eqnarray}
where $a_0$ and $a$ are the unstrained and strained (bilayer) primitive lattice constant. A similar quantity can be defined for germanene on graphene. As shown in Table I, we focus on the supercell models that induce a strain of less than 3$\%$. The bond angles $\theta$ and buckling distance $\Delta$ in silicene and germanene will be slightly affected by the strain as shown in Table I. In freestanding silicene or germanene, the bond angle $\theta$ is uniform. With the presence of a substrate, the sublattice symmetry is slightly broken in the Si or Ge layer, hence the bond angles exhibit a small variation of a few degrees. Shown in Table I are the average values.
\begin{figure}
\includegraphics[width=2.9in]{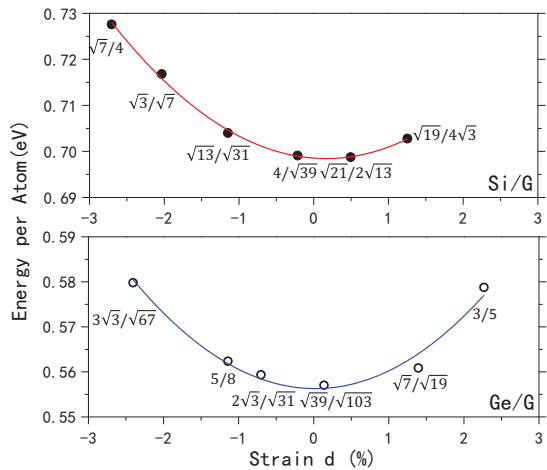}
\caption{(Color online). Energy per Si or Ge with reference to the bulk value for silicene or germanene on graphene obtained using different supercell models in Table I. The results are plotted as a function of strain in the layer.}
\end{figure}
The vdW interaction between the layers requires special attention. In order to address the interlayer interaction, we define a binding energy ($E_{b}$, per Si atom) in the Si/G bilayer as
$E_{b}$ = $(E_{\text{Si/G}}-E_{\text{G}}-E_{\text{Si}})/N_{\text{Si}}$,
where $E_{\text{Si/G}}$, $E_{\text{G}}$, and $E_{\text{Si}}$ are the total energies in the same supercell for Si/G, monolayer graphene, and monolayer silicene, respectively, and $N_{\text{Si}}$ is the number of Si atoms in this supercell. This binding energy for the Si($\sqrt{7}$)/G(4) bilayer is evaluated by various exchange-correlation functionals, and the results as a function of the layer separation are shown in Fig. 2. (For the purpose of examining interlayer interaction, the energy data presented in Fig.~2 are calculated at graphene's in-plane lattice constant. A mesh of 400 $k$ points in the primitive graphene BZ and a plane-wave energy cutoff of 800 eV are used.) Except for the Perdew-Burke-Ernzerhof (PBE) generalized gradient approximation\cite{pbe} that fails to create any binding between the layers, all other functionals (LDA, PBE-vdW-D2, and PBE-vdW-DF\cite{vdw,sevdw}) predict energy minima at an interlayer separation around 3.3$-$3.6 $\text{\AA}$. The LDA gives an energy lowering of 58 meV per Si atom due to the interlayer interaction, while the functionals with explicit vdW corrections significantly increase the energy gain with the PBE-vdW-DF functional yielding the largest gain of 121 meV per Si atom.  Since we are mostly concerned with relative energies and the electronic structure in the present work, and the variation in the interlayer separation is not expected to significantly affect the results, in the following we will report LDA results at an interlayer separation of 3.3 $\text{\AA}$ for the simplicity of the calculations unless otherwise noted. In comparison, the LDA gives an average binding energy of 63 meV per Ge atom in supercells (Table I) due to the interaction with the graphene substrate.
\begin{figure}
\includegraphics[width=2.9in]{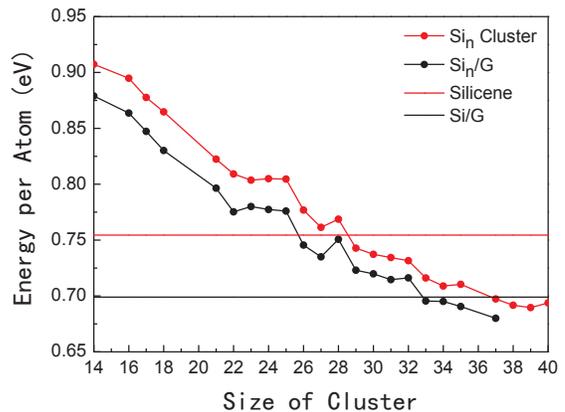}
\caption{(Color online). The relative energy per Si atom (compared with the bulk value) in monolayer silicene, Si($\sqrt{21}$)/G($2\sqrt{13}$), Si$_n$ cluster, Si$_n$, and Si$_n$/G.}
\end{figure}
\begin{figure*}
\includegraphics[width=6.5in]{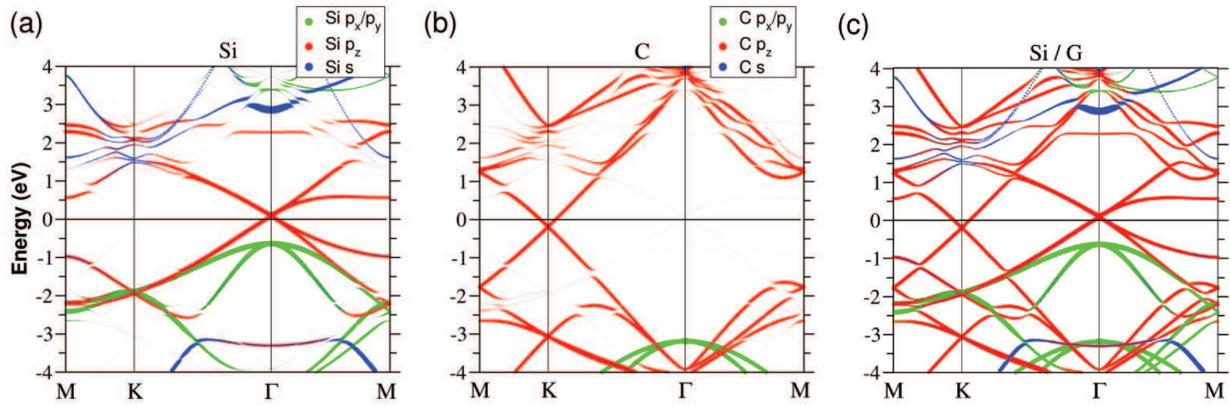}
\caption{(Color online). Band structure of Si($\sqrt{3}$)/G($\sqrt{7}$): (a) the projected states on Si are highlighted; (b) the projected states on C are highlighted; and (c) the projected bands in (a) and (b) are combined. The substrate-induced gap is about 26 meV for Si ($\Gamma$) and 2 meV for graphene ($K$), respectively.}
\end{figure*}
The energetics of the silicene overlayer can be addressed by examining the energy per Si atom defined as
\begin{eqnarray}
E_{c}=\frac{(E_{\text{Si/G}}-E_{\text{G}})}{N_{\text{Si}}}-\mu_{\text{Si}},
\end{eqnarray}
where $\mu_{\text{Si}}$ is the chemical potential set to the energy per atom of bulk Si.

A similar expression can be defined for the germanene overlayer. The calculated energies per Si (Ge) atom using different supercell combinations in Table I are plotted against the strain in Fig. 3, with the minimum being around zero strain as expected. The positive energy values indicate that the 2D structure is higher in energy than the 3D diamond structure. Among the structures we have considered for Si/G, Si(4)/G($\sqrt{39}$) and Si($\sqrt{21}$)/G($2\sqrt{13}$) are the two structures with the smallest strain ($-$0.2\% and 0.5\%, respectively) and the lowest energy. For Ge/G, Ge($\sqrt{39}$)/G($\sqrt{103}$) has the smallest strain (0.1\%) and the lowest energy. The energy difference per atom between different supercell models is smaller than the thermal energy at room temperature, indicating that multiple phases of different crystalline orientation could coexist at room temperature.
\begin{figure}
\includegraphics[width=2.9in]{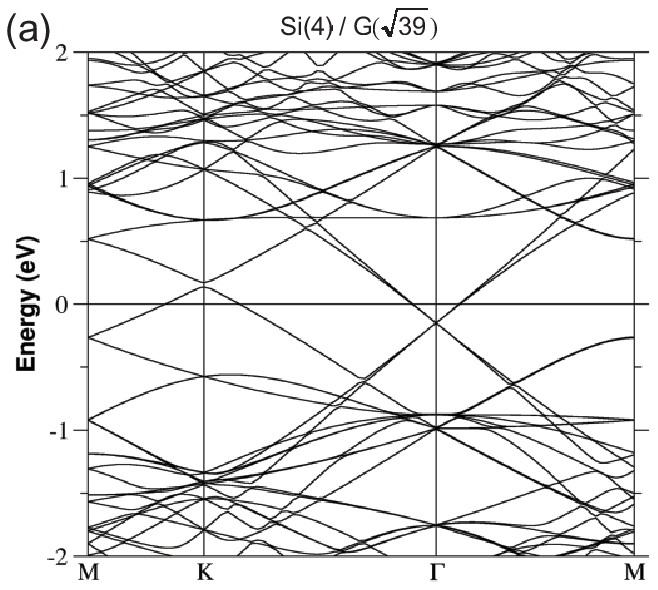}
\includegraphics[width=2.9in]{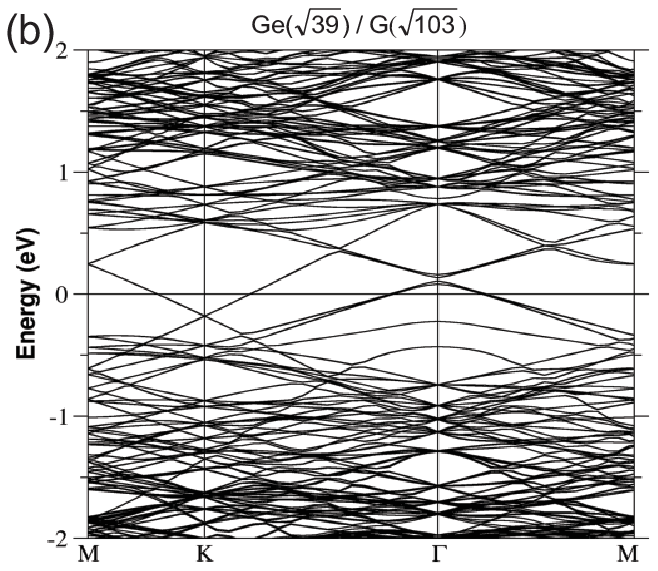}
\caption{Band structure of (a) Si(4)/G($\sqrt{39}$), the Dirac point of silicene (graphene) is located at $K$ ($\Gamma$), and (b) Ge($\sqrt{39}$)/G($\sqrt{103}$), the Dirac point of germanene (graphene) is located at $\Gamma$ ($K$).}
\end{figure}
We have also calculated the phonon modes at the $\Gamma$ point of Si/G and did not find any imaginary frequency, which is a necessary condition for a stable LB honeycomb structure of Si and Ge.\cite{ciraci09} The sliding barrier is only 0.4 meV per Si atom, implying that silicene could slide easily on graphene due to the weak vdW interaction.

\begin{figure}
\includegraphics[width=2.9in]{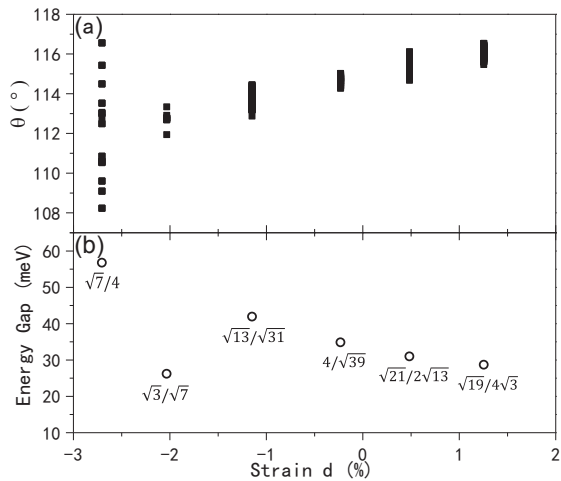}
\caption{The angle distribution of silicene bonds over unit cell and the energy gap of Si/G as a function of local strain.}
\end{figure}
In order to study the role of graphene as a substrate in the synthesis of silicene and to access the stability of 2D silicene in comparison with 3D Si clusters, we have calculated the energies of free and adsorbed Si$_n$ clusters ranging from $n=14$ to 40 Si atoms and compared them with that of silicene. The energies per Si atom compared with that of bulk Si are shown in Fig. 4. The initial structures of Si$_n$ clusters are obtained from the Cambridge cluster database (CCD) by S. Yoo $et$ $al$.\cite{data1,data2,data3,data4,data5} For Si$_n$ on graphene, a 6$\times$6 graphene supercell is used.  Comparing the energy per atom in free Si$_n$ clusters (red dots) and in freestanding silicene (red line), we conclude that the small free Si$_n$ cluster is less energetically favorable than silicene when $n\leq28$. With the presence of the graphene substrate, the Si$_n$/G curve (black dots) is lower than that of Si/G (black line) when $n\geq 33$. Therefore, based on the energetics results, the graphene substrate increases the possibility of growing 2D silicene over forming 3D Si clusters. The bonding in Ge clusters is expected to be similar to that in Si clusters, hence we expect a similar conclusion for Ge clusters on graphene.

For multiple layers of silicene on graphene, we find that the structure is stable in bilayer, but becomes highly distorted structures above three layers. This result indicates the difficulty of growing more than two silicene layers. Similar results are found for Ge/G.

\section{Electronic structure}

Two-dimensional honeycomb structures, buckled or not, exhibit a symmetry between the sublattices and therefore have a linear energy dispersion in the vicinity of the Dirac points at $K$ and $K^{'}$. Figure 5 shows the projected band structures of Si($\sqrt{3}$)/G($\sqrt{7}$) on Si and C atoms, as well the whole energy spectrum. The contributions from $p_x/p_y$ (degenerate), $p_z$, and $s$ orbitals are presented in green, red, and blue, respectively. The projected-band structure on Si resembles that of freestanding silicene, where the $p_z$ states (red) are responsible for the $\pi$ bonds in the vicinity of the Fermi level. For Si($\sqrt{3}$)/G($\sqrt{7}$), the Dirac point of silicene is mapped to $\Gamma$ [see Fig.~5(a)], while the Dirac point of graphene stays at $K$ [see Fig.~5(b)]. The Fermi level crosses the two Dirac cones of silicene and graphene, generating a small amount of electron transfer from silicene to graphene. The amount of charge transfer is about 2$\times10^{12}$ cm$^{-2}$ or $5\times10^{-4}$ electrons per C atom and corresponds to a Fermi-level shift of about 0.1 eV in silicene. It is noted that the $s/p_x/p_y$ states of the Si atoms are much closer to the Fermi level than those in graphene, indicating a weaker $sp^2$ configuration in silicene with a mixing of the $sp^3$ characteristics. The characteristic of Dirac fermions are preserved in all slab models accompanied with a small amount of charge transfer from silicene or germanene to graphene, this can be clearly seen in Fig.~6 by plotting the band structure of lowest-energy slab models for Si/G and Ge/G we studied, Si(4)/G($\sqrt{39}$) and Ge($\sqrt{39}$)/G($\sqrt{103}$).

The graphene substrate introduces an inhomogeneous potential that breaks the sublattice symmetry of silicene. For a freestanding silicene monolayer with the LB structure, the bond angles between a Si atom and its nearest neighbors are uniform. For the supported layer, the bond angles have a variation, hence the sublattice symmetry is broken, and a gap is opened. In Fig.~7, we plot the distributions of the bond angles (black squares) over the unit cell for different slab models, as well as the corresponding gaps at the Dirac points of silicene. For Si($\sqrt{7}$)/G(4), a gap as large as 57 meV is obtained in silicene as a result of the substrate interaction, while the overall electronic structure is not altered significantly. For most other configurations we have considered, the gaps are in the range of 25$-$40 meV. In comparison, only a very small gap, in general less than 5 meV, is opened in graphene bands.

As discussed in previous studies, the SOC lifts the degeneracy between the upper and lower bands at the Dirac point and opens a gap. In germanene, the SOC effect is about ten times larger than that in silicene.\cite{yao11} We find that for Si/G, the SOC splitting is less than 2 meV, which is quite small compared to the substrate-induced gap (see Fig.~7). For Ge/G, as in the case of Ge($\sqrt{12}$)/G($\sqrt{31}$), the 48-meV gap of germanene at the Dirac point is reduced to 23 meV after adding the SOC, suggesting that the interplay between the substrate and SOC effects could be used to tune the band gap in these bilayer systems.

\section{Conclusion}

To conclude, we have shown that, by first-principles calculations, it is possible to synthesize silicene and germanene on the graphene substrate without destroying its characteristics of the Dirac-fermion-like linear dispersion around Dirac points, due to the weak van der Waals interlayer interaction. In addition, multiple phases of single crystalline silicene or germanene with different orientations could coexist at room temperature based on our energetics analysis. The substrate breaks the sublattice symmetry in silicene and germanene and induces a gap at the Dirac point. For silicene on graphene, the gap could be as large as 57 meV. For germanene on graphene, the gap created by the substrate effect is of the same order as that induced by the SOC effect. The interplay between the substrate and SOC effects could be used for further band-gap manipulations. Our fundamental study of the electronic structure and energetics of these silicene/graphene and germanene/graphene bilayers may provide important insight for other two-dimensional van der Waals heterostructures.

\section{Acknowledgments}

We thank Professor Yong Zhang for helpful discussions. This work is supported by the Academia Sinica and the National Science Council under Grants No. 101-2119-M-002-008 and 99-2112-M001-034-MY3. M.Y.C. acknowledges partial support from the U.S. Department of Energy (Grant No. DE-FG02-97ER45632).

\bibliographystyle{apsrevM}
\ifx\mcitethebibliography\mciteundefinedmacro
\PackageError{apsrevM.bst}{mciteplus.sty has not been loaded}
{This bibstyle requires the use of the mciteplus package.}\fi

\end{document}